\documentstyle[12pt]{article}
\begin{document}
\begin{titlepage}
\vspace{35mm}
\begin{center}
{\Large \bf Multifractality and Multiscaling in two dimensional 
fragmentation} 
\end{center}

\vskip 10mm

\begin{center}
{\bf M K Hassan$^{\rm a b}$ and G J Rodgers$^{\rm a}$}
\end{center}

\vskip2mm

\begin{center}
\footnotesize
\begin{tabular}{ll}
a\ &Department of Physics, Brunel University, Uxbridge, Middlesex UB8
      3PH, UK \\
b\ &Department of Physics,
Shahjalal Science and Technology University, Sylhet, Bangladesh
\end{tabular}
\end{center}

\vspace{20mm}

\begin{center}
{\bf Abstract}
\end{center}
\noindent

\vspace{20mm} 

\noindent
We consider two models (A and B) which can describe both two dimensional 
fragmentation and stochastic fractals. Model A exhibits multifractality 
on a  unique  support when describing a fragmentation process and 
on one of infinitely many  possible supports when describing 
stochastic fractals. Model B obeys simple scaling.  

\vspace{8mm}


\vspace{3mm}

\noindent
Keywords: \hspace{4mm}Fragmentation, stochastic, fractal, multifractal 

\vspace{15mm}

\vspace{3mm}

\noindent
\footnotesize {E-mail: Kamrul.Hassan@brunel.ac.uk and 
G.J.Rodgers@brunel.ac.uk}

\end{titlepage}

\newpage
\vspace{5mm}

Fragmentation is an irreversible kinetic process 
 in which a collection of fragments are sequentially broken. There have 
been a 
number of different analytical approaches to the problem of the 
kinetics of
fragmentation of one dimensional particles. These have included using the 
maximum entropy method [1], using statistical and combinatorial 
arguments [2,3] and using a 
kinetic equation. It is the kinetic equation approach, developed by 
Filippov [4] after it's original proposal by Kolmogorov [5], 
that has provided most of our theoretical understanding. This has 
resulted in numerous exact and explicit solutions for the 
particle size distribution function [6,7], in addition to scaling 
solutions [8,9].

In one dimension the size or mass of the particles is the 
only dynamical quantity of interest. However, there  has been a recent 
theoretical 
interest [10-12,20] in the kinetics of fragmentation of multidimensional 
objects. This is  motivated by a desire to move away from 
characterising particles solely  by their volume or equivalently, mass 
and towards 
an understanding of the physical role played by shape in the 
fragmentation of particles. In reality, particles are identified by their 
shape.

The studies of fragmentation phenomenon in two dimensions [10-12, 20] have 
revealed unexpectedly rich patterns with interesting and novel 
statistics.
We attempt to clarify the origin of this behaviour 
 and invoke  the idea of multifractality to characterise such pattern.
We find that  some fragmentation rules  do not have a 
single measure support instead they have an infinite number. Each  
measure yields different spectrum of 
exponents to characterize the system. 
We also consider a second model which exhibits simple scaling and we seek 
to explain the difference between these models.

\noindent
The obvious extension of the rate equation to two dimensions 
[10-13]] is \begin{eqnarray}
{{\partial f(x,y;t)}\over{\partial t}} = 
-f(x,y;t)\int_0^xdx_1\int_0^ydy_1F(x_{1},y_1,x,y)+  \nonumber \\
       s\int_x^{\infty}dx_1\int_y^{\infty}dy_1f(x_1,y_1;t)F(x,y,x_1,y_1) 
\end{eqnarray}  
where $f(x,y;t)$ is the concentration of particles of sides $x$ 
and $y$ at time $t$ and $s=1,2,3$ or $4$. 
 $F(x_1,y_1,x,y)$ describes the rate with which objects having sides 
$x$ and $y$ break to produce fragments of sides $x_1$, $x-x_1$ and 
$y_1$, $y-y_1$.

The two integrals over the two variables implies that the two orthogonal 
cracks are placed on an objects such that the cracks are equal and 
parallel to the sides. Evidently they will produce four new fragments at 
each time step to describe fragmentation process ($s=4$).  However, we can 
choose  $s=1,2,3$ which simply implies that $s$ fragments are kept and $(4-s)$
fragments are removed at each time step. This process creates a  
stochastic fractals  [14,15]] at long times. In this letter, we consider 
our two variables $x$ and $y$ to be lengths. This is in contrast to [12], 
where one variable was associated with energy and the other with mass.

We choose to study a homogeneous rate kernel [12],

\begin{equation}
F(x_1,y_1,x,y)=x^{\beta_1}y^{\beta_2}
\end{equation}

This model (called A in the abstract) describes a system in which particles 
are selected for fragmentation with a rate determined by their 
area 
and shape. The relative  importance of area and shape is 
determined by $\beta_1$ and $\beta_2$.
Once a particles has been chosen for fragmentation, products 
of any area and shape are equally likely so that the daughter distribution is 
Poisson. Substituting this choice of kernel into the rate equation yields

\begin{equation}
{{\partial f(x,y,;t)}\over{\partial t}} = 
-x^{\beta_1+1}y^{\beta_2+1}f(x,y;t)+
s\int_x^{\infty}\int_y^{\infty}dx_1dy_1x_1^{\beta_1}y_1^{\beta_2}f(x_1,y_1;t)
\end{equation}

\noindent
We now define the moments of the probability distribution 
function $f(x,y;t)$ as

\begin{equation}
M_{m,n}(t) = \int_0^{\infty}\int_0^{\infty}dxdyx^{m-1}y^{n-1}f(x,y;t)
\end{equation}
We can then obtain a rate 
equation for the moments,

\begin{equation}
{{\partial M_{m,n}(t)}\over{\partial t}}=({s\over{mn}}-1)M_{m+\beta_1+1,n+
\beta_2+1}(t)
\end{equation}
An interesting feature of the above equation is that there are
infinitely many conserved (time independent) moments. Moments 
$M_{m,n}(t)$ that are conserved satisfy $mn=s$. This simply reflects the 
fact that fragments with a given area 
can have an infinite number of different shapes.

Using Charlesby's method, the moment equation can be iterated to 
get all 
the derivatives of the moments [11,12]. These can then be substituted into a 
Taylor series expansion of 
$M_{m,n}(t)$ about $t=0$ to give

\begin{equation}
M_{m,n}(t) = \ _{2}F_2(a_1,a_2;b_1,b_2;-t)
\end{equation}

\noindent
$_2F_2$ is a generalised hypergeometric function [15], where 

\begin{equation}
a_{1/2} = {m \over{2(\beta_1 +1)}}+{n\over{2(\beta_2+1)}}
\mp\sqrt{{(m(\beta_2+1)-n(\beta_1+1))^2+4s
(\beta_1+1)(\beta_2+1)}\over{4(\beta_1+1)^2(\beta_2+1)^2}}
\end{equation} 
and
\begin{equation}
b_1= {m\over{(\beta_1+1)}}
\end{equation}
\begin{equation}
b_2 = {n\over{(\beta_2+1)}}
\end{equation}
We are only interested in the long time behaviour of the moments.
The asymptotic 
expansion of the generalised hypergeometric function for large time $t$ gives
\begin{equation} 
M_{m,n}(t)\approx {{\Gamma (b_1) \Gamma (b_2) \Gamma(a_1-a_2)} 
\over{\Gamma (a_2) \Gamma(b_1-a_1)\Gamma(b_2-a_1)}}t^{-a_1} 
\end{equation}
For $s=4$ the conserved moments can be written as  
$M_{m^{\ast},{{4}\over{m^{\ast}}}}(t)$ where, 
$m^{\ast}$ is any number. We will only consider $m^\ast$ positive. 
Obviously, $M_{2,2}(t)$ can be identifyed as the 
area of the system. For convenience, we 
choose $\beta_1=\beta_2=\beta$ since it does not change the physics of 
the mechanism under investigation.

We may choose to 
associate each hidden conserved quantity of the $s=4$ 
process with the set of points in ${\cal R}^2$. This space can 
then be subdivided into boxes of size

\begin{equation}
\delta_{m\ast} = 
\sqrt{{{M_{m^\ast,{{4}\over{m^\ast}}}(t)}\over{M_{1,1}(t)}})}.
\end{equation}
such that $\mu_i(\delta_{m^\ast})$ denotes the measure within the 
$i^{th}$ box that depends on the choice of $s$. This choice of 
$\delta_{m^\ast}$ ensures that for a pure fragmentation process ($s=4$), 
we recover the complete set of points in ${\cal R}^2$. We 
now express the moment $M_{m,1}(t)$ in terms of $\delta_{m^\ast}$ 
\begin{equation}
M_{m,1}(\delta_{m^\ast}) \sim 
\delta_{m^\ast}^{-\gamma(m^\ast)(\sqrt{(m-1)^2+4s}-m-1)} \end{equation}
where 
\begin{equation}
\gamma(m^\ast)={{2}\over{(m^\ast+{{4}\over{m^\ast}})-\sqrt{(m^\ast 
-{{4}\over{m^\ast}})^2+4s}-2+\sqrt{4s}}}
\end{equation}
When $m=1$, the exponent of equation (13) gives the 
Hausdorff-Besicovitch dimension $D_f$.  
Note that $\gamma(m^\ast)= \gamma(4/m^\ast)$ and that in the limit 
$m^\ast \rightarrow \infty$, when $s>1$, $\gamma(m^\ast) \rightarrow 
1/(\sqrt{s}-1)$. We 
 also see that for $s=4$, $\gamma(m^\ast)=1$ and $D_f=2$, independent of 
$m^\ast$, as we ensured with our choice of $\delta_{m^\ast}$.
However, for $1<s<4$  there exist infinitely many 
supports for different values of $m^\ast$ and each is subdevided 
by a corresponding $\delta_{m^\ast}$. We also can write the $d-$measure 
of the weighted box number $M_{m,1}(t)$ as 
\begin{equation}
M_{m,1}(d, \delta_{m^\ast})=\sum_i 
\mu_i^{k(m)}\delta_{m^\ast}^d = 
N(k(m),\delta_{m^\ast)})\delta_{m\ast}^d.
\end{equation}
Where $ N(k(m),\delta_{m^\ast}^d)$ is the $k(m)^{th}$ moment of the 
measure such that $N(0,\delta_{m^\ast})$ is the number of boxes 
require to cover the support of dimension $d=D_f$. Hence we can write the 
weighted box number as
\begin{equation}
N(k(m),\delta_{m^\ast}^d)= 
\sum_i\mu_i^{k(m)} \sim 
\delta_{m^\ast}^{-\tau(k(m))}
\end{equation}
We thus see that $M_{m,1}(\delta_{m^\ast})$ can be partioned into boxes 
of sides $\delta_{m^\ast}$ such that the probabilities $\{\mu_i\}$ 
are normalised if we let
\begin{equation}
m=1+(s-1)k
\end{equation}
when $\delta_{m^\ast}\rightarrow 0$ we require that the measure 
$M_{m,1}(d,\delta_{m^\ast})$ tends to a finite value. This occurs when 
$d=\tau(k)$. Combining (13),(15) and (17) gives the mass exponent 
$\tau(k)$ as
\begin{equation}
\tau(k)=\gamma(m^\ast)\{\sqrt{(s-1)^2k^2+4s}-(s-1)k-2\}
\end{equation}
\noindent
This expression meets some  essential  requirement; namely $\tau(0)$ is the 
dimension of 
the support and $\tau(1)=0$. 

We thus see that there exist a spectrum of mass exponents $\tau(k)$ that 
characterise the distribution of the particle size distribution.
The mass exponent is nonlinear which indicates the existence of a 
fractal subset for each support whether or not the support itself is 
fractal. 
To find this
 fractal subset we use the usual Legendre transform of the independent variables 
$\tau$ and $k$ to the independent variable $\alpha$ and $f(\alpha(k))$;
\begin{equation}
\alpha(k)=-{{d\tau(k)}\over{dk}}
\end{equation}
and 
\begin{equation}
f(\alpha(k))=k\alpha(k)+\tau(k)
\end{equation}
These relations yield 
\begin{equation}
\alpha(k)=\gamma(m^\ast){{(s-1)\sqrt{(s-1)^2k^2+4s}-(s-1)^2k}\over{\sqrt{(s-1)^2k^2+4s}}}
\end{equation}
and 
\begin{equation}
f(\alpha(k))=\gamma(m^\ast){{4s-2\sqrt{(s-1)^2k^2+4s}}\over{\sqrt{(s-1^2k^2+4s}}}
\end{equation}
It is interesting to note that all the quantities of interest, 
$\gamma(m^\ast)$, $\tau(k)$ and $f(\alpha(k))$ are independent of $\beta$ 
when $\beta_1=\beta_2=\beta$. 
 When $\beta_1 \neq \beta_2$ 
all the quantities depend on both $\beta_1$ and $\beta_2$ so 
the analogous expressions are much more complex. However, the basic 
picture is unchanged. For all $\beta_1$ and $\beta_2$ the $f-\alpha$ 
spectrum obeys a simple scaling relationship with respect to $\gamma$, 
namely $f(\alpha)=\gamma h(\alpha/\gamma)$. In figure $1$ the $f-\alpha$ 
spectrum is plotted for three different values of $m^\ast$.

Physically, the $f(\alpha(k))$ versus $\alpha$ curve simply suggest 
the existence of intertwined fractal subsets describing the measure.
We find that when fragments are removed from the system at each time step 
 there 
exist a range of fractal dimension $1.4641 \leq D_f(m^\ast) \leq 2$ and 
there are an infinitely manny $f-\alpha$ spectra for each $D_f$. All the 
$D_f(m^\ast)$ compete on equal footings to be the support on which the 
measure can distributed in a given realisation. This reflects that in 
addition to the entropy from the location of the fractal subset there is 
another source of entropy from $D_f(m^\ast)$ which is absent in the pure 
fragmmentation ($s=4$) process.

As an aside, let us mention the connection between these models 
for $1<s<4$ and 
those of random sequential adsorption (rsa) (see [17] for a recent review). 
One can imagine that at each time step, the fragmentation event is a  
 deposition 
in which $4-s$ fragments are deposited on the substrate and play no further 
part in the 
kinetics and $s$ regions of the substrate survive for future deposition. 
The difference between true rsa and our system is that in our system 
deposition can only take place in rather restricted set of position 
also as time proceeds the deposited particles get smaller in size. 
However, in the long time the pattern created looks rather like that 
created by the deposition of a mixture of sizes of ractangles. This 
is the $2-d$ variant of the $1-d$ system of mixture deposition 
studied by [18]. The $s=2$ version of this system studied in [19] the 
context of deposition of needles. In rsa, one of the interesting 
observables is that system reaches a jamming limit at long time 
which is less than random close packing. However, in this case we find 
instead of having a  jamming limit the number density shows power law 
behaviour with non-trivial exponent.
In general, the jamming limit can be found using the following relation
\begin{equation}
\theta(t \rightarrow \infty)=1-\int_0^\infty dx\int_0^\infty dy f(x,y;t)
\end{equation}
This gives
\begin{equation}
1-\theta(t \rightarrow \infty) \sim  t^{\sqrt{s}-1}
\end{equation}
To understand the role 
played by the dimension and shape we consider now a different model ie
\begin{equation}
F(x_1,y_1;x,y)=x^{\beta_1}y^{\beta_2}\delta(2x_1-x)\delta(2y_1-y)
\end{equation}
Substitute this into the rate equation to get,
\begin{equation}
{{\partial f(x,y;t)}\over{\partial 
t}}=-{1\over{2^2}}x^{\beta_1}y^{\beta_2}f(x,y;t)+2^{\beta_1+\beta_2}sf(2x,2y;t)
\end{equation}
And the rate equation for the moments
\begin{equation}
{{\partial M_{m,n}(t)}\over{\partial 
t}}=-({1\over{2^2}}-{s\over{2^{m+n}}})M_{m+\beta_1,n+\beta_2}(t)
\end{equation}
This again gives infinitely many hidden conserved quantity. As before the 
condition for the conserved dynamical quantity can be 
obtained from the requirement of time independent moments
\begin{equation}
m+n=2+{{{\ln}s}\over{{\ln}2}}
\end{equation}
This linear relation between $m^,$s and $n^,$s implies asymptotic power 
law decay of the moments with linear exponent in $m^,$s and $n^,$s in time:  
\begin{equation}
M_{m,n}(t) \sim A(m,n)t^{-\alpha(m+n)} 
\end{equation}
Substitute this into the rate equation for the moments to give 
a difference equation 
\begin{equation}
\alpha(m+n+\beta_1+\beta_2)=\alpha(m+n)+1
\end{equation}
Iterating the above difference equation and using the appropriate 
boundary condition gives
\begin{equation}
\alpha(m+n)= {{{{m+n-(2+{{{\ln}s}\over{{\ln}2}})}}}\over{(\beta_1+\beta_2)}}
\end{equation}
This gives us the power law decay of  all the moments and allows us to 
show that the average number of fragments $<N(t)>$ and the average area 
are related by
\begin{equation}
<N(t)> \sim <xy>^{-{{{\ln}s}\over{2{\ln}2}}}
\end{equation}
Consequently, the fractal dimension for $s<4$ is given by
\begin{equation}
D_f={{{\ln}s}\over{{\ln}2}}
\end{equation}
and we have $f(\alpha)=D_f$ and $ \tau(k)=D_f(1-k)$ (self-similar). In 
short, model B exhibits simple scaling.

\vspace{3mm}

These results give us the opportunity to ask why one needs an 
infinite number of independent exponents to characterise the scaling 
relations in model A while model B only exhibits scaling. 
To find the answer we need to go back to the nature of the models 
themselves and search for the things we lost in moving from the model A 
to the model B. In the model A we had stochastic 
homogenieity which implies the fragmentation of an object 
possesses ergodic probability distribution. In this model, two 
orthogonal cracks are placed independently parallel to the sides ie they 
can pass through 
any points in Euclidean space. While the  
model B describe two orthogonal cracks are allowed to 
place only at the middle of the objects to produce successfully four equal 
sized fragments. It implies that the size is no longer intrigued with 
shape ie shape is determined by the initial condition. Note that it is 
one of the infinitely many possibility of the former model. Thus if there 
is a mixture of particles of different size and shape, and if any fragments 
are 
equally likely to be picked, in the second model, once a fragments with 
definite shape 
is picked for fragmentation that will only produce of that shape. Thus it is 
the broken ergodicity in shape that causes the absence of multiscaling.

In conclusion, we have studied the two different models to understand the 
fragmentation phenomenon when there are more than one dynamical 
variable. We found  significantly different behaviour between the two 
models, although both models have an infinite number of conserved moments. 
Interestingly, these  models also helps us to explain  the occurence of 
multifractality in fragmenting systems, which 
is not yet fully understood.

 \newpage 
\begin{flushleft}
{\large \bf Acknowledgement} 

\vskip5mm  

MKH would like to thank the CVCP for an ORS award. 
\end{flushleft}

\vspace{8mm}

\begin{flushleft}
{\large \bf  References}
 
1. Englemen R 1991 J. Phys: condensed matter {\bf 3} 1019 

2. Mark H and Simha R 1940 Trans. Faraday Soc. {\bf 35} 611 

3. Montroll E W and Simha R 1940 J. Chem. Phys. {\bf 8} 721  
 
4. Filippov A F 1961 Theory Probab. Appl. {\bf 4} 275  

5. Kolmogorov A N 1941 Dan USSR (Doklady) {\bf 31} 99  

6. Ziff R M and McGrady E D 1985 J. Phys. A {\bf 18} 3027 

7. McGrady E D and Ziff R M 1987 Phys. Rev Lett {\bf 58} 892  

8. Williams M M R 1990 Aero. Sci. and Technol. {\bf 12} 538  

9. Cheng Z and Redner S 1990 J Phys A {\bf 23} 1233  

10. Rodgers G J and Hassan M K 1994 Physical Rev. E {\bf 50} 3458 

11. Krapivsky P L and Ben-Naim E 1994 Phys Rev E {\bf 50} 3508 

12. Boyer D Tarjus G and Viot P 1995 Phys Rev E {\bf 50} 1043  

13. Singh P and Hassan M K 1996 Phys. Rev. E in press 

14.  Krapivsky P L Ben-Naim E 1994 Physics Letters A {\bf 196} 168  

15. Hassan M K and Rodgers G J  Physics Letters A  {\bf 208} 95 

16. Luke Y L 1969 The special functions and their approximation {\bf 1} 
( New York, Academic Press)  

17. Feder J, 1988 Fractals (New York, Plenum). 

18. Evans J W 1991 Rev. Mod. Phys. {\bf 65} 1281  

19. Krapivsky P L 1992 J. Stat. Phys. {\bf 69} 135  

20. Tarjus G and Viot P 1991 Phys. Rev. Lett. {\bf 67} 1875 

\end{flushleft}
\newpage

{\bf Figure}

\vspace{5mm}

1.	Three of the $f-\alpha$ spectra for model A when $s=3$ and 
$\beta_1=\beta_2$. The three curves are for $m^\ast =2,4$ and $\infty$.
 
\end{document}